# The Persistent Effect of Famine on Present-Day China: Evidence from the Billionaires[*]

Pramod Kumar Sur[†]

Asian Growth Research Institute (AGI) and Osaka University

Masaru Sasaki

Osaka University and IZA

**Abstract**

More than half a century has passed since the Great Chinese Famine (1959–1961), and China has transformed from a poor, underdeveloped country to the world's leading emerging economy. Does the effect of the famine persist today? To explore this question, we combine historical data on province-level famine exposure with contemporary data on individual wealth. To better understand if the relationship is causal, we simultaneously account for the well-known historical evidence on the selection effect arising for those who survive the famine and those born during this period, as well as the issue of endogeneity on the exposure of a province to the famine. We find robust evidence showing that famine exposure has had a considerable negative effect on the contemporary wealth of individuals born during this period. Together, the evidence suggests that the famine had an adverse effect on wealth, and it is even present among the wealthiest cohort of individuals in present-day China.

[*] This paper was previously circulated as "Famine and Wealth Inequality." We are very grateful to Charles Yuji Horioka, Hiroyuki Kasahara, Hongyong Zhang, Katsuya Takii, Lex Zhao and seminar and conference participants at Asian and Australian Society of Labor Economics (AASLE), Asian Growth Research Institute (AGI), European Association of Labor Economists (EALE), Japanese Economic Association (JEA), Hitotsubashi University, Kansai Development Microeconomics Seminar (KDME), Kobe University, Labor Economics Conference, and Trans Pacific Labor Seminar (TPLS). All remaining mistakes are our own.

[†] Corresponding author. Email: pramodsur@gmail.com.

# 1. Introduction

Between 1959 and 1961, China experienced the deadliest famine in history. Over the course of three years, millions of people literally starved to death (Dikotter, 2010). Although mortality estimates vary widely, about 16.5–45 million people perished, making it one of the deadliest events in the 20th century.[1]

However, China is now quite different from how it was the past. Since opening its economy in 1978, China has achieved rapid economic growth and transformed from a poor, underdeveloped country to an upper-middle income country and the world's second-largest economy. China's GDP growth rate has averaged almost 10% a year, and more than 850 million people have been lifted out of poverty.[2] Between 1978 and 2015, China's per-capita income multiplied more than eight times, and its share of world GDP increased from only about 3% in 1978 to 20% in 2015 (Piketty et al., 2019).[3]

Does the famine have any economic impact that persists today? If so, who are the vulnerable groups or cohorts of populations? Addressing these questions is an issue of great importance, as millions of survivors who were exposed to the famine are still alive today. Moreover, a better understanding of how different groups or cohorts of the population have benefited (or not) from the enormous economic growth in China is needed. This issue is relevant from a point of view of fairness, as previous studies have found that China's rapid growth has benefited people disproportionately, and inequality has increased dramatically since the late 1970s (Piketty et al., 2019). Finally, understanding and addressing these questions are fundamental and relevant in general, as the existential threat of famine and the lingering effects of those that have passed still exist outside of China and are ever relevant. This paper addresses these questions and examines the effect of exposure to China's Great Famine on individual wealth outcomes more than half a century afterward.

---
[1] Estimations range from 16.5 million (Coale, 1981) to 45 million (Dikotter, 2010).
[2] World Bank estimates. https://www.worldbank.org/en/country/china/overview#1
[3] Annual per-adult national income rose from less than 6,500 yuan in 1978 to over 57,800 yuan in 2015 (PPP estimates).

The primary concern when studying the effects of extreme events, such as famine, is the issue of a selection effect arising for those born during this period and those who survived the famine.[4] For example, in China's case, along with a sharp rise in the death rate, the birth rate during the famine period (1959–1961) was the lowest in two decades between 1950 and 1970 (see Appendix Figure A1). Among those who died during this period, mortality rates were particularly high among children and the elderly (Ashton et al., 1984; Spence, 1990). Moreover, earlier studies have argued that when famine is severe and mortality rates are high, survivors are more likely to comprise "select" individuals who have naturally stronger constitutions. Those from the lower parts of the distribution are more likely to die. Therefore, survivors are usually from the top of the distribution of important characteristics, such as physical resilience, income and access to nourishment (Deaton, 2008; Gørgens et al., 2012). In addition to the selection effect, there is a further concern of endogeneity because exposure of a province to famine itself may be endogenous to other factors. If such characteristics are systematically different and likely associated with later economic outcomes, comparing the mean of the distribution of survivors with the mean of the control group's distribution could bias the true impact of famine.

We employ three strategies to address simultaneously the issue of selection effect and endogeneity concerns in this paper. First, to address the issue of positive selection on survival, we estimate the effect of famine exposure on the upper quantiles of the distribution of wealth outcomes. This method is proposed by Meng and Qian (2009), assuming that these estimates will more accurately reflect the actual effect of exposure to famine because survivors are from the top of the distribution of outcomes. For this purpose, we consider individual-level wealth data from the Hurun Report. The report, otherwise known as the "Hurun China Rich List," ranks wealthy individuals in China and publishes their current wealth every year. This dataset is particularly useful for our estimation, as these individuals are at the top of the current wealth distribution.[5]

---

[4] Several earlier studies have pointed out the potential concerns of attenuation bias caused by the selection for survival as well as endogeneity concern (Friedman, 1982; Bozzoli et al., 2009; Gørgens et al., 2012).

Our second strategy is to empirically investigate whether there are any systematic selection issues related to survival or those born during this period in our sample of individuals. Examining this is essential because one could argue that the concept of the upper quantile of the distribution of outcomes suggested by Meng and Qian (2009) can be arbitrary. It is difficult to empirically establish where to draw a line or set a boundary such as above the mean, above the median, 4th quartile, or 99th percentile. Besides, we need to be certain that the dataset we are considering does not have any selection bias. We perform a simple test to examine the selection issue in this paper. We calculate the number of individuals born in each province each year (cohort size) from our dataset and consider it as our outcome variable. We then compare exposed cohort sizes with unexposed cohort sizes for each province where famine intensity was high or low to check for any systematic difference of exposure to the famine on the number of individuals born in a province before, during and after the famine. Based on this, we show that there are no systematic selection issues related to survival or those born during this period in our sample of individuals. To the best of our knowledge, this is the first study to empirically examine and test the selection effect for studying the impact of famines.

After confirming that our dataset does not have any systematic selection effect, our third and final strategy is to address endogeneity, considering that exposure of a province to the famine itself may be endogenous to other unobserved factors. As our initial approach, we employ a difference-in-difference (DD) strategy by exploiting province-by-cohort level variation in exposure to the famine as a quasi-experiment. Combining contemporary individual-level wealth data with historical data on famine severity by province, we question whether individuals' exposure to the famine at different periods of their life has any effect on wealth that could persist today. Our hypothesis builds on well-established insights from the medical and health literature suggesting that exposure to a shock, especially during a person's early life, can have persistent and profound impacts later (see Almond and Currie, 2011, for a review).

---

[5] This dataset, not unique to us, has been widely used recently to study wealth concentration at the very top in China; see Xie and Jin (2015) and Piketty et al. (2019).

We find that the famine accounted for a substantial decline in wealth for individuals born during this period. In fact, our estimation shows that, on average, a 1 percentage point increase in exposure to the famine led to a 1.4–2.0% decline in wealth. In terms of magnitude, the famine we examine caused an average reduction in wealth of 29.6–52.0% for individuals born during this period. We also provide a variety of alternative estimates suggesting that our results are robust to the measurement of wealth in a specific year, restricting individual cohorts to 15, 10 and 5 years before and after the famine, and different clustering choices. Finally, we present our DD results through an event study design and visually show that there is no systemic trend in our control sample, and the parallel trend assumption is likely to be satisfied.

After establishing that the famine in China accounted for a substantial decline in wealth for individuals born during this period, we next turn to the task of addressing whether other events may have affected our estimation. As we noted earlier, China also experienced additional events during the period on which we focus, including a sharp decline in the birth rate during the famine and a sharp increase in the birth rate immediately after the famine. Furthermore, it also suffered from a decade of Cultural Revolution (1966–1976) after the famine, a result of which was the closure of secondary and tertiary educational institutions during the late 1960s and early 1970s. We formally test whether these events affected the wealth difference uncovered in our analysis.

We perform two different exercises to address whether other events affected our estimation. First, we conduct falsification tests by assigning pseudo-treatment. We perform this exercise to check whether or not the treatment has a causal effect on the outcome. In particular, we pseudo-treat individuals born three years immediately before (1956–1958) and after the famine (1962–1964) as the placebo-affected cohort for the before and after cohorts, respectively, and perform falsification tests for each group. Additionally, we restrict each of our cohorts to those born within six years immediately before and after the famine and perform similar exercises. In each case, we find that the wealth of individuals born before and after the famine period is unaffected by the province-level famine intensity.

Second, we test our findings through an alternate measure of famine severity, the excess death rate. We perform this exercise to confirm whether any bias associated with our measurement of famine exposure drives the results. In our baseline analysis, we consider the average cohort size loss, which is commonly used, to retrospectively measure exposure during the famine period. However, bias could occur because severe fluctuation in birth rates (as we mentioned before) may reflect endogenous fertility decisions during this extreme period. Similarly, the average loss of cohort size would fail to capture the mortality rates of adults and the elderly. We use the province-level excess death rate during this period based on statistics published by the National Bureau of Statistics of China (1999) as a direct and alternative measure of famine severity and obtain results consistent with our main findings.

Notably, our dataset consists of only currently wealthy individuals from the Hurun Report. We are aware that the individuals we consider in our estimation are the wealthiest and belong to the top of the distribution of wealth in the Chinese population. As we demonstrate in this paper, we use this dataset to simultaneously examine and address the issue of endogeneity and the selection effect to provide causal evidence. We further show that a robust and sizable negative effect of famine is even present among these wealthiest cohorts of individuals. These caveats should be kept in mind while interpreting the findings from our results.

The paper speaks to several diverse literatures. First, the magnitude of the effect and the cohort of individuals we study contribute to the broader literature on understanding the severity as well as the persistent economic impact of "fragile environments" such as famine (Neugebauer et al., 1999; Ravelli et al., 1999; Brown et al., 2000; Hulshoff Pol et al., 2000; Chen and Zhou, 2007; Meng and Qian, 2009; Almond et al., 2010; Neelsen and Stratmann, 2011; and Dercon and Porter, 2014). The Great Chinese Famine appears to be an extreme example in terms of providing insight into the effect of exposure to famine on wealth. However, famines are not unique to China. Famine has affected hundreds of millions of people alive today in developing as well as developed countries at some point during their lifetime. Conflict and fragility are also persistent and still seen in many developing countries today. Given the importance of understanding the effect of such an environment, we build on the previous works in three ways. First, our improved estimates suggest that individuals born in such environments can suffer *very* significant adverse effects that could persist for an extended time. Second, we study Chinese billionaires (millionaires in terms of US dollars) and show that the adverse effect remains even among these *"strongest and wealthiest"* cohorts of individuals. Finally, and importantly, our findings provide some suggestive evidence showing that decades of rapid economic development may not mitigate the adverse impact.

Second, our paper also contributes to the growing literature on understanding the causes of uneven growth of income and wealth in China today. In particular, many papers indicate that inequality has risen rapidly in China recently. For example, Piketty et al. (2019) suggest that China's inequality levels were close to those of the Nordic countries in the late 1970s but are now approaching those of the US. Notably, the biggest increase took place between the mid-1980s and mid-2000s. We build on this work and provide a potential mechanism for the rapid rise in inequality in China since the famine cohorts entered the labor market during this period. Nonetheless, our focus on the historical determinants of difference in wealth should not imply that other factors are unimportant. A number of existing studies have shown the importance of determinants such as escalating housing prices, differential saving, capital accumulation and changes in the legal system regarding property contributing to wealth inequality in China (e.g., Li and Wan, 2015; Knight et al., 2017; Piketty et al., 2019). As we demonstrate here, a potential historical legacy affecting the current unequal distribution of wealth remains in China even today.

Finally, this study also broadly relates to the literature that provides a methodological improvement for the examination of the long-term effect of exposure to famine, such as Meng and Qian (2009) and Gørgens et al. (2012). We build on this work by simultaneously addressing and testing the issue of the selection effect and provide an improved estimate. The method we propose and use to examine the selection bias is intuitive and easy to implement. This method can also be applied to other contexts similar to famine, where researchers have similar priors regarding the pattern of selection, such as war, the refugee crisis and the recent COVID-19 pandemic.

The remainder of the paper is structured as follows. Section 2 provides a brief background of the Great Chinese Famine. Section 3 describes the historical and contemporary data used in the empirical analysis. Section 4 presents the results, and Section 5 presents conclusions.

**2. The Great Chinese Famine**

In this section, we provide a brief discussion about the Great Famine. For a detailed discussion along with the historical background of the famine, please see Meng et al. (2015).

There is controversy among social scientists and historians over the exact timing of the Great Famine. The Chinese Government, however, officially defines it to be three years between 1959 and 1961 when the mortality rates were the highest. The mortality figures are also controversial and vary widely. It is estimated that between 16.5 and 45 million people died during this period (Coale, 1981; Dikotter, 2010).

The extant literature on the Great Famine also debates the factors that primarily led to what ultimately became a nationwide calamity. According to the official explanation provided by the Chinese government, the main reason for the famine was the fall in output due to bad weather. However, recent studies have provided evidence that the fall in output was also partly due to bad government policies such as the diversion of resources away from agriculture to industrialization, as well as weakened worker incentives. One strand of research argues in favor of the food availability decline (FAD) hypothesis that most associate with the Great Leap Forward (GLF) and the collectivization of agriculture that began in 1958 (Lin, 1990; Yao, 1999). The GLF and the collectivization of agriculture resulted in a drastic decline in grain production in 1959, and this continued for the next two years before coming to a halt in 1962. Another belief is that food wastage from communal dining during the GLF was partly responsible for the famine (Chang and Wen, 1997).

By contrast, an alternative strand of research focuses on factors that led to entitlement failure. For example, it is believed that politically zealous and career-concerned officials exaggerated grain production figures to create a good impression of the success of collectivization and exported rice to the urban population and that this intensified famine in rural areas (Lin and Yang, 2000; Kung and Chen, 2011). Furthermore, recent findings by Kasahara and Li (2019) suggest some evidence that grain exports used to repay loans from the Soviet Union and the import of industrial equipment to promote the GLF further intensified the famine in China.

Regardless of the causes, both the urban and rural populations in China experienced an increase in mortality rates during the famine period. However, the rural rate in 1960 was 2.5 times the pre-famine rate. Urban residents fared better but were not spared, with death rates at their peak in 1960 being 80% above their pre-famine level (National Bureau of Statistics of China 1999). The famine intensity also varied widely by province. For example, Anhui and Sichuan were among the worst-affected provinces, whereas Heilongjiang and Inner Mongolia were among the least affected. Nevertheless, the famine caused widespread excess deaths in China. From the perspective of the excess number of deaths, the Great Chinese Famine outstrips any famine in recorded history.

## 3. Data Sources and their Description

### 3.1 Historical Data

The historical data on province-level famine intensity we use for our main analysis comes from the 1990 China Population Census. The census reports on a 1% sample of the universe of China's population. To construct the province-level famine intensity, we first calculate the average cohort size for the three years before (i.e., 1956–1958) and after (i.e., 1962–1964) the famine for each province. We then calculate the average cohort size during the famine (i.e., 1959–1961) for each province. We measure the famine intensity as the percentage decrease from the average pre- and post-famine cohort size to the famine's cohort size.[6] This measure essentially captures the percentage of missing people in the famine cohort in each province. In Figure 1, we report the famine intensity of provinces across China for our estimation based on our calculations from the

---

[6] At the time, Chongqing was part of Sichuan Province. Therefore, we consider that the famine intensity of Chongqing equals that of Sichuan Province.

1990 China Population Census.[7]

The alternative measure of famine intensity that we use in this paper comes from the National Bureau of Statistics of China (1999).[8] This reports the death rate by province each year between 1949 and 1998. From the report, we construct the famine intensity replicating to the extent possible the census methods. In particular, we first calculate the average death rate (per thousand) for the three years before (1956–1958) and after (1962–1964) the famine for each province. We then calculate the average death rate (per thousand) during the famine (1959–1961) for each province and measure the excess death rate as the percentage increase from the average pre- and post-famine death rate to the death rate during the famine.[9] In Figure A2 in the Appendix, we present the famine intensity of each province in China based on the excess death rate.

*3.2 Contemporary Data*

We combine the historical data on famine intensity with the wealth data from the Hurun Report. The report lists wealthy individuals and family members in China and reports their current wealth every year.[10] The Hurun Report is quite similar to the Forbes World's Billionaires list that ranks wealthiest individuals globally based on their current wealth in US dollars. The Hurun Report ranks individuals or family members who hold a minimum wealth of 2 billion Chinese yuan.[11] We collect the dataset of the universe of Chinese populations listed in the report between 2015 and 2017. The full sample includes 6,058 entries, with 2,570 unique individuals and joint family members. First, we restrict our sample to single individuals' wealth holdings and exclude joint wealth holdings listed in the report. We do this because it is challenging to estimate each individual's share of wealth accurately in joint wealth holdings. Second, we exclude entries for which the birth year is not available because we need to know the year of birth to examine the effect of exposure to the famine. Finally, of the three years we consider in this paper, the 2016 report only contains each individual's province of birth. As the province of birth is our primary variable used to estimate the variation in the exposure to famine, we restrict our sample to individuals listed in the 2016 report and construct the panel of individuals (2015–2017) using that report. Our final dataset consists of 2,948 entries, with 1,049 unique individuals. Table B1 in the Appendix provides a detailed description of the complete dataset and the dataset used in this study.

---

[7] We exclude Tibet, as our sample includes no individuals born in Tibet.
[8] The same dataset can also be obtained from Lin and Yang (2000).
[9] Notably, Chongqing and Hainan were part of Sichuan and Guangdong provinces during the famine period, respectively. We thus consider the famine intensity of Chongqing and Hainan equal to that of Sichuan and Guangdong provinces, respectively.
[10] http://www.hurun.net/EN/Home/
[11] It equal to about 295 million US dollars in 2015 and 300 million US dollars in 2016 and 2017, respectively.

We have two outcome variables. The first and key outcome is the total wealth held by an individual listed in the report. We construct this variable by calculating the natural logarithm of wealth held by each individual each year in US dollars. Figure A3 in the Appendix reports the average wealth of individuals based on their year of birth and average wealth by famine intensity of their province of birth. The blue and red lines represent the average wealth of individuals born in provinces where the famine intensity was above and below the average, respectively. As we can see, Figure A3 depicts a parallel trend in the wealth holding of individuals based on famine intensity, except for the extremes.

Our second outcome is the cohort size of individuals by province and year of birth. We construct this variable to determine whether there are any systematic selection issues related to survival or those born during this period in our sample of individuals. To construct the cohort size, we calculate the number of unique individuals born in each province each year from our dataset.[12] The oldest and the youngest individuals in our sample were born in 1935 and 1986, respectively.

---

[12] The value is zero if there are no individuals born in a province in a year.

Furthermore, no individual was born in years 1936, 1938 and 1985. Therefore, the cohort size variable constitutes a province-level balanced panel dataset for 49 years.

## 4. Analysis

### 4.1 Selection Effect

Before estimating the effect of famine on wealth, we first test the selection issues for survival among those born during the famine period in our dataset using the following equation:

$$Cohort_{py} = \alpha + \lambda_1(FamineIntensity_p \times Famineperiod_y) + \lambda_2(FamineIntensity_p \times Beforeperiod_y) + \delta Province_p + \mu Year_y + \varepsilon_{py} \quad (1)$$

where $Cohort_{py}$ is the cohort size of individuals in province $p$ and year $y$. $FamineIntensity_p$ is the province-level average decrease in cohort size during the famine period that we calculate from the 1990 China Population Census. We include province fixed effects ($Province_p$) and year fixed effects ($Year_y$) to control for the fact that provinces may be systematically different from each other and nationwide common shocks, respectively. $\varepsilon_{py}$ is a random, idiosyncratic error term.

We divide the timing of a cohort's exposure to the famine into two groups to examine the selection effect of birth and the selection effect of survival. We construct a before-famine-period dummy variable ($Beforeperiod_y$) that takes a value of one if the year is before 1959 and zero otherwise. We create this variable to examine any selection effect of survival only. Second, we construct a famine-period dummy variable ($Famineperiod_y$) that takes a value of one if the year is between 1959 and 1961 and zero otherwise to examine any selection effect of birth and survival during the famine period. Our identification strategy is a generalized DD estimation where the principal treatment variables are the interactions between the percentages of the excess mortality rate in a province ($FamineIntensity_p$) with our two groups of cohorts.

We present the estimates for the selection effect of famine in Table 1. Column (1) of Table 1 indicates that there is a negative relationship between exposure to the famine on cohort size in a province before and during the famine. However, the coefficients are not statistically different from zero. The results are similar when we examine the effect of a smaller time-window in Columns (2) and (3), considering the fact there are not many individuals in our sample at the two extremes. Our results are also robust to excluding samples where the outcome is zero (see Table B2 in the Appendix). Finally, we present our findings through an event study specification in Figure A4 in the Appendix. The black line reports the main effect with a 95% confidence interval represented by the dotted gray line. As can be seen, the findings are similar and robust to alternative specifications. Overall, these results confirm that there are no systematic selection issues in our sample of individuals.

*4.2 Main Effect*

After confirming that our dataset does not have any systematic selection issues, we now move in this section to examine the effect of the Great Famine on wealth. To account for the endogeneity concern, we exploit province-by-cohort level variation in famine intensity as a natural experiment. Our identification strategy is a generalized DD estimation where the principal treatment variables are the interactions between the excess mortality rate, in percentages, during the famine period (famine intensity) with a dummy variable identifying those born before and during the famine. In particular, the proposed estimates of the average treatment effect are given by $\lambda_1$ and $\lambda_2$ in the following baseline province of birth, birth year, and year of ranking fixed effects equation:

$$\ln(Wealth_{iypt}) = \alpha + \lambda_1(BornDuring_{iy} \times FamineIntensity_p) + \lambda_2(BornBefore_{iy} \times FamineIntensity_p)$$
$$+ \mu Birthyear_y + \delta Birthprov_p + \eta RankingYear_t + \varepsilon_{iypt} \qquad (2)$$

where $\ln(Wealth_{ipyt})$ is the natural log of the total amount of wealth held by individual $i$ born in province $p$ and year $y$ for ranking year $t$ in US dollars. $FamineIntensity_p$ is the province-level average decrease in cohort size during the famine period in comparison with the general trend (pre and post). $Birthprov_p$ is the province of birth fixed effect, controlling for the fact that provinces may differ systematically. $Birthyear_y$ is the birth year fixed effect, controlling for nationwide common shocks. $RankingYear_t$ is the individual's ranking year (in the Hurun Report) fixed effect, controlling for the common change in wealth over time.

It is well known in the medical and health literature that the timing of exposure to famine or a health shock could have heterogeneous effects on later economic outcomes (see Almond and Currie, 2011 for a review). In particular, exposure to the famine among individuals born during the famine (in-utero exposure) have different effects than for individuals born before the famine started. Based on these insights, we divide individuals into two groups, depending on when they were exposed. $BornDuring_{iy}$ is a dummy variable that takes a value of one if individual $i$ was born during the famine (1959–1961) and zero otherwise. $BornBefore_{iy}$ is a dummy variable that takes the value one if individual $i$ was born before the famine (1958 or before) and zero otherwise. Finally, $\varepsilon_{ipyt}$ is a random, idiosyncratic error term. For our baseline analysis, we cluster the standard errors at the birth year, allowing error terms to be correlated across individuals within the same birth cohorts across provinces. We do this considering the general fact that wealth increases with age. Given that we use a panel of individual data covering 2015–2017, we estimate a random-effects model, as the individual effects are not correlated with the treatment.

We report the baseline results in Table 2. Each column is a separate regression. Columns 1–3 report the main results of estimating equation (2) with alternative specifications. Our first DD estimate of $\lambda_1$ is reported in the first row, with the estimates showing that $\lambda_1$ is negative at the 1% level of significance in every specification. After controlling for the various fixed effects in column 3, the results suggest that a 1 percentage point increase in famine intensity leads to a 1.4% decrease in wealth on average. In terms of magnitude, the difference in wealth based on exposure to famine is quite significant. In our calculation, the province-level famine intensity was 37.28% on average during this period. This implies that at its mean, the famine caused a decrease in wealth of about 52%. Our second DD estimate for $\lambda_2$ is reported in the third row of Table 2. As we can observe, the coefficients are small and positive, but statistically insignificant from zero in every specification. The results show no significant effect on individuals born before the famine and exposed to different famine intensity on wealth. In summary, our overall findings suggest that in-utero exposure (born during the famine) to famine has a significant negative effect on wealth, but that otherwise, the general effect of exposure to famine is minimal. We also find that the effect of famine is large and persists in China for at least five decades.

As discussed earlier, we considered the individuals listed in the 2016 Hurun Report and constructed panel data over 2015–2017, as the province of birth is only available in the 2016 report. There may be some concern that our estimations may be misleading, as the main results reported in Table 2 do not account for individuals listed in the 2015 and 2017 reports but not included in the 2016 report. If they are somehow correlated with famine or disproportionately related at the province level, these excluded individuals in our estimation could produce a biased result. To check the robustness of our findings, we estimate equation (2) considering individuals' wealth listed in the 2016 report only. We report the results in Table B3 with various alternative specifications. As we can see, the results are statistically and quantitatively similar to the baseline results reported in Table 2.

As Figure A3 in the Appendix illustrates, there is a wide variation in individual wealth at the two extremes. One of the reasons is that the sample sizes each year (the number of individuals born each year and listed in the report) is small. Another concern could be that the older and younger cohorts may differ from the middle-aged category on various dimensions, which we may not capture in our estimation. If the unobserved differences are somehow systematically correlated in our estimation, it may bias the results. Furthermore, in our baseline estimation, we consider the full sample of individuals to examine the effect of the general exposure to famine (born before the famine variable) on any differences in reported wealth. There might be some concern that the effect of exposure to famine may differ by age. In particular, medical research and recent research in economics suggest that infants (below five years old) are most vulnerable to such events (Currie and Almond, 2011).

To test this, we estimate equation (2) with alternative specifications and report the results in Table 3. For comparison, we present the baseline estimation in column 1, similar to column 3 in Table 2. In column 2, we estimate the coefficients by restricting our sample to individuals born 15 years before and after the famine (i.e., between 1944 and 1976). Furthermore, in column 3, we limit our sample to individuals born 10 years before and after the famine (i.e., between 1949 and 1971) and perform our regression analysis. Finally, in column 4, we restrict our sample to individuals born five years before and after the famine (i.e., between 1954 and 1966) and perform our analysis. The results shown in columns 2–4 are relatively consistent and quantitatively similar to our baseline results reported in column 1. In summary, we find that the wide variation in individual wealth, especially at the two extremes (the very old and very young cohorts), is not driving our main findings. Furthermore, we find that childhood exposure to famine (the first 5, 10 and 15 years of life) has no statistically significant effect on wealth holdings, which is consistent with our main findings. Moreover, in line with the 'fetal origins hypothesis' (Barker, 1990;

Almond and Currie, 2011), we reveal a significant negative effect on individuals' wealth holdings if they were born during the Great Famine.

Up to this point, we have examined the combined effect of exposure to famine on wealth. Next, we present the year-by-year effect of exposure to famine through event study graphs. We show the graph for the whole period in Figure A5 in the Appendix, and the effect between 1950 and 1970 in Figure 2 for clear visualization. Panel A reports the year-by-year effect for the whole sample (2015–2017). In Panel B, we further report the event study graph for the 2016 sample only. The black line shows the main impact, with a 95% confidence interval represented by the dotted gray line. As we observe, each famine year's coefficient is negative and significantly different from zero. Furthermore, the estimated coefficients of individuals born after the famine period do not show any specific pattern confirming that the parallel trend assumption is more likely to hold. Unfortunately, we do not have any information regarding an individual's actual date and month of birth, which would have allowed us to examine the mechanism in more detail. We refer this question to future research.

*4.3. Falsification Exercise through Pseudo-treatment*

In Section 4.2, we establish that the famine in China accounted for a substantial decline in wealth for individuals born during this period. We also provide a variety of alternative analyses showing that our results are robust to a specific year of measurement of wealth, restricting the sample to 15, 10 and 5 years before and after the famine, and estimating through event-study design. In this section, we turn to the task of addressing the concerns of other events affecting our estimation by conducting falsification tests, where we assign pseudo-treatment.

In our falsification test, we focus on the older cohort already born before the start of the famine (the before cohort) and the younger cohort born after the famine (the after cohort). In our specification, we pseudo-treat individuals born three years immediately before (1956–1958) and after the famine (1962–1964) as the placebo-affected cohort for the before and after cohorts, respectively, and perform a falsification test for each group.[13] Furthermore, because we revealed some heterogeneity in the very oldest and very youngest cohorts, we restrict our samples to those born immediately before or after the famine (within six years before and after) and check the robustness of our findings. We present the results in Table 4. Columns 1 and 2, and columns 3 and 4 perform the falsification exercise on the before and after cohorts with various specifications, respectively. From Table 4, we find that the differences in wealth holdings between the before and after cohorts are similar and not significantly different from zero. In other words, the falsification test suggests that the wealth of individuals born before and after the famine is unaffected by the province-level famine intensity. The falsification exercise also lends some confidence that our treatment (i.e., exposure to the famine) is more likely to have a causal effect on our outcome.

### *4.4 Inference through Alternative Choices of Clustering*

In our baseline estimation in Table 2, we adjust standard errors for clustering at the birth cohort level, allowing error terms to be correlated across individuals within the same birth cohorts across provinces. In this section, we explore to what extent our baseline inference presented in Table 2 is affected by alternative choices of clustering.

We present the results with various alternative choices of clustering in Table B4 in the Appendix. In Column 1, we report robust standard errors without clustering. In Column 2, we re-estimate our baseline specifications, where we allow error terms to be correlated across individuals born in the same province. In Column 3, we report baseline specifications re-estimated by implementing the two-way clustering to allow error terms to be correlated within the province of birth and within the cohort of birth. As we can see, our results are robust to these alternative choices of clustering.

---

[13] There are multiple ways of assigning pseudo-treatments, such as random assignment to some years, assigning to multiple years beyond the three years that we assign. We do it this way for simplicity and convenience of interpretation.

*4.5. Estimation through Alternative Measure of Famine Intensity*

Another potential confounding factor in our main findings could be the measurement of famine intensity. We noted in Section 3 that we construct the famine intensity measure using the missing birth cohorts from the 1990 census. However, studies on the Great Chinese Famine and Figure A1 in the Appendix highlight a sharp decline in the birth rate during the famine period. One could argue that fertility decisions are endogenous, especially during a severe catastrophic situation, such as the Great Chinese Famine. The missing birth cohort also fails to capture the mortality rates of adults and the elderly. Additionally, there is evidence of a sudden increase in the birth rate immediately after the famine, which could inflate the missing birth cohort during the famine period. Therefore, our measure of famine intensity may be severely biased. This section provides an additional check for our main findings by estimating equation (2) using the excess death rate as a direct and additional measure of famine intensity derived from the National Bureau of Statistics of China (1999).

We present the full sample results in Table 5 and for the 2016 sample in Table B5 in the Appendix. As we observe, these estimates are similar and statistically consistent with our baseline results reported in Table 2. According to the regression estimates in column 3 of Table 5, a 1 percentage point increase in famine intensity (based on excess death rates) leads to, on average, about a 1.9% decrease in wealth. According to this estimate, the famine caused a reduction of 29.6% of wealth on average (the average excess death rate in the sample is 14.8%). Comparing this with our earlier estimates, the magnitude is a bit smaller. One potential reason for the lower magnitude might be the systematic under-reporting of official death statistics during this period.

## 5. Concluding Remarks

This paper examines the effect of exposure to China's Great Famine on wealth more than half a century afterward. To estimate the impact, we combine contemporary individual-level wealth data with historical data on famine severity in China. To better understand if the relationship is causal, we simultaneously account for the well-known historical evidence on the selection effect arising for those who survive the famine and those born during this period, as well as the issue of endogeneity on the exposure of a province to the famine. We provide strong evidence suggesting that exposure to famine has a large negative effect on the wealth of individuals born during this period. Additionally, we show that the effect is present after more than half a century, and it is even persistent among the *wealthiest cohort of individuals* in China today.

As with any study that relies on natural experiments, our results can be considered local to our context. However, we believe this context is of particular interest as we examine the economic consequences of a famine of which millions of survivors are still living today. In particular, investigating the historical persistence of famine among 'literally billionaires' who belong to the *strongest and wealthiest* cohort of individuals in present-day China, and showing that decades of rapid economic development may not mitigate the adverse impact merits important implications for understanding how profound is the impact of famine in China. Moreover, the findings from this study may also have implications for understanding the role of historical legacy on the current unequal distribution of wealth in China.

These findings may also be relevant to our understanding of the long-term consequences of famine in other countries, as famines are not unique to China alone. Famine has affected hundreds of millions of people alive today in developing and developed countries. Similarly, famines are also persistent and still seen in many developing and middle-income countries today. For example, as of 2018, drought has affected more than 22 million people in East Africa, and at least 15 million people were going hungry.[14] Additionally, 6.8 million people experienced extreme hunger in Yemen, 13.5 million people were in need of assistance in Syria, and 5 million refugees had fled to other countries.[15][16] Thus, viewed more broadly, this study provides some insight into understanding the persistent effect in countries that have faced such events in the past and the future consequences for those currently facing these events. Although only a piece of anecdotal evidence, Ukraine and Kazakhstan are the most unequal countries in the world today (based on the Gini wealth index) and are countries that have also faced severe famines in the past.[17] Further research on these issues is necessary to provide rigorous analysis at the macro level.

---

[14] https://www.oxfam.org.uk/what-we-do/emergency-response/east-africa-food-crisis. Accessed October 2, 2018.
[15] https://www.oxfam.org.uk/what-we-do/emergency-response/yemen-crisis. Accessed October 2, 2018.
[16] https://www.oxfam.org.uk/what-we-do/emergency-response/syria-crisis. Accessed October 2, 2018.
[17] According to the Credit Suisse Global Databook 2018, Ukraine and Kazakhstan rank first and second based on the Gini wealth index, respectively. https://www.credit-suisse.com/about-us/en/reports-research/global-wealth-report.html. Accessed October 2, 2018.

Figure 1: Province-level Famine Intensity

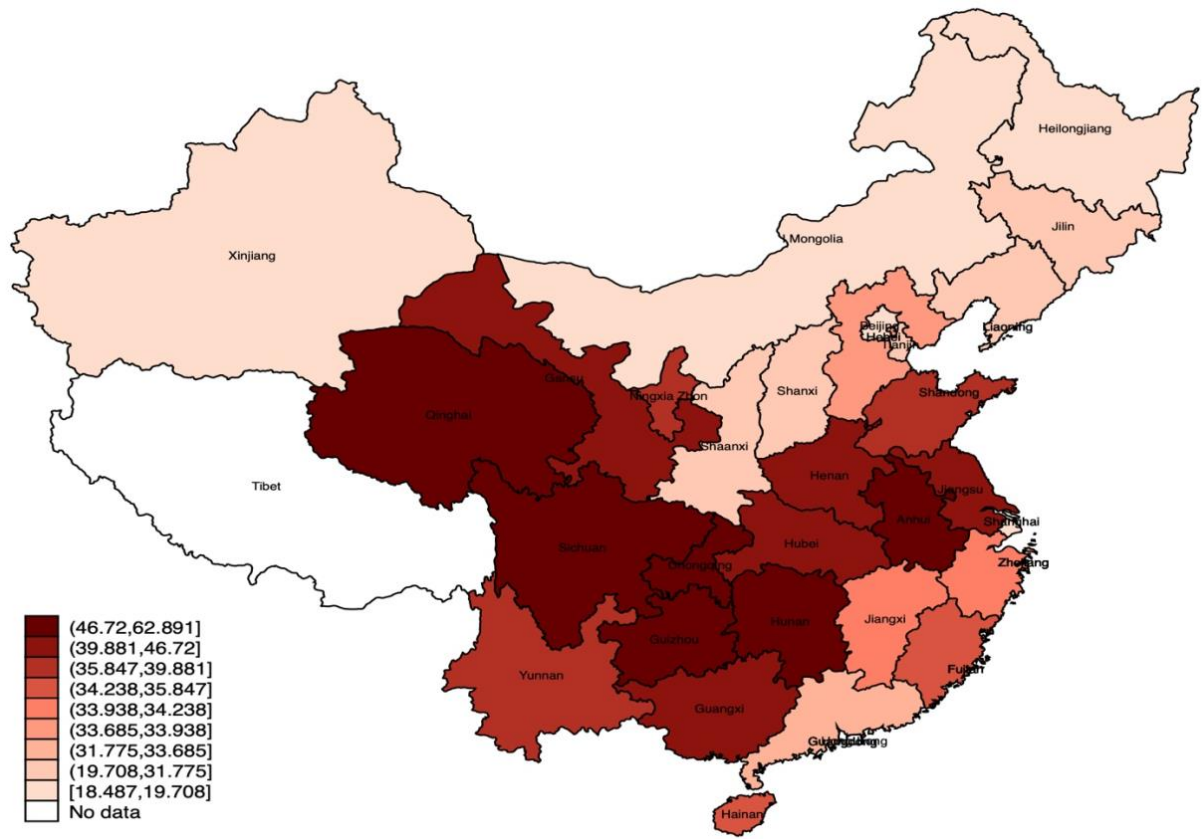

*Notes:* The figure presents the province-level variation in famine intensity measured by the percentage of missing people in the famine cohort in comparison with three years prior (i.e., 1956–58) and after (i.e., 1962–64) the famine. Darker shades indicate larger share of missing people during the famine.

Figure 2: Event Study (1950-1970)

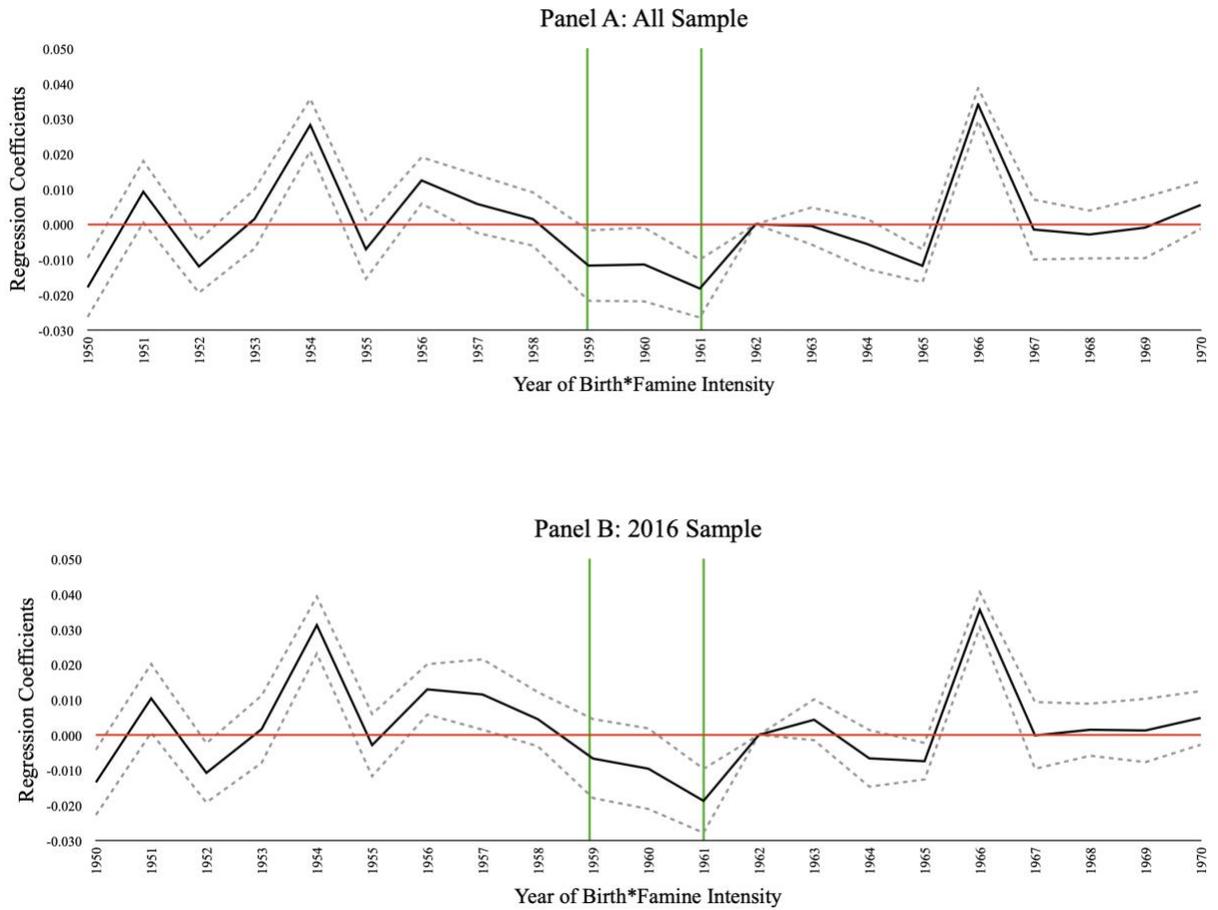

*Notes:* The figure presents the regression coefficients of the year-by-year effect of exposure to the famine between 1950-70 through event study graphs. Panel A reports the event study graph for the whole sample (2015-17). Panel B reports the event study graph for 2016 sample only. The black line reports the main impact with a 95% confidence interval presented in the dotted grey line.

Table 1: Impact of Famine on Cohort of Birth by Province

|  | Dependent Variable: Cohort Size | | |
| --- | --- | --- | --- |
|  | All Years (1) | 1940-1980 (2) | 1950-1970 (3) |
| Famine Intensity*Famine Period | -0.004 | -0.006 | -0.013 |
|  | (0.006) | (0.006) | (0.010) |
| Famine Intensity*Before the Famine Period | -0.003 | -0.005 | -0.009 |
|  | (0.003) | (0.004) | (0.009) |
| Province FE | YES | YES | YES |
| Year FE | YES | YES | YES |
| Observations | 1,519 | 1,271 | 651 |

*Notes:* Data are constructed from 2016 Hurun Report. The unit of observation is a province. *Cohort Size* is the number of unique individuals born in each province each year in our sample. *Famine Intensity* is the province-level average decrease in cohort size during the famine calculated from the 1990 China Population Census. *Famine Period* is the year between 1959 and 1961. *Before the Famine Period* is the year before 1959. Robust standard errors adjusted for clustering within province. *** Significant at the 1% level. ** Significant at the 5% level. * Significant at the 10% level.

Table 2: Impact of Famine on Wealth

|  | Dependent Variable: Wealth (Log) | | |
| --- | --- | --- | --- |
|  | (1) | (2) | (3) |
| Famine Intensity*Born during the Famine | -0.0125*** | -0.0124*** | -0.0142*** |
|  | (0.003) | (0.003) | (0.004) |
| Famine Intensity*Born before the Famine | 0.003 | 0.003 | 0.002 |
|  | (0.005) | (0.005) | (0.005) |
| Birth Year FE | Yes | Yes | Yes |
| Province of Birth FE | No | No | Yes |
| Ranking Year FE | No | Yes | Yes |
| N | 2948 | 2948 | 2948 |

*Notes:* Data are from 2015-2017 Hurun Report. The unit of observation is an individual. *Wealth (Log)* is the natural log of the total amount of wealth held by individuals in the US dollars. *Famine Intensity* is the province-level average decrease in cohort size during the famine calculated from the 1990 China Population Census. *Born during the Famine* is a dummy variable for individuals born between 1959 and 1961. *Born before the Famine* is a dummy variable for individuals born before 1959. Columns 1 and 2 control for province-level famine intensity but are not shown here. Robust standard errors adjusted for clustering within the birth year. *** Significant at the 1% level. ** Significant at the 5% level. * Significant at the 10% level.

Table 3: Sample Restriction to Specific Years

|  | Dependent Variable: Wealth (Log) | | | |
|---|---|---|---|---|
|  | Full Sample | 15 years before and after | 10 years before and after | 5 year before and after |
|  | (1) | (2) | (3) | (4) |
| Famine Intensity*Born during the Famine | -0.0142*** | -0.0137*** | -0.0126*** | -0.0136** |
|  | (0.004) | (0.003) | (0.004) | (0.006) |
| Famine Intensity*Born before the Famine | 0.002 | 0.003 | 0.004 | 0.009 |
|  | (0.005) | (0.005) | (0.005) | (0.008) |
| Other Controls | Yes | Yes | Yes | Yes |
| N | 2948 | 2809 | 2513 | 1629 |

*Notes:* Data are from 2015-2017 Hurun Report. The unit of observation is an individual. *Wealth (Log)* is the natural log of the total amount of wealth held by individuals in the US dollars. *Famine Intensity* is the province-level average decrease in cohort size during the famine calculated from the 1990 China Population Census. *Born during the Famine* is a dummy variable for individuals born between 1959 and 1961. *Born before the Famine* is a dummy variable for individuals born before 1959. Other control variables include birth year fixed effects, ranking year fixed effects and province of birth fixed effects. Robust standard errors adjusted for clustering within birth year. *** Significant at the 1% level. ** Significant at the 5% level. * Significant at the 10% level.

Table 4: Falsification Exercise through Pseudo-treatment

|  | Dependent Variable: Wealth (Log) | | | |
| --- | --- | --- | --- | --- |
|  | Born Before the Famine (All Sample) | Born Between 1953–58 | Born After the Famine (All Sample) | Born Between 1962–1967 |
|  | (1) | (2) | (3) | (4) |
| Born Between 1956–58*Famine Intensity | -0.000 | -0.010 |  |  |
|  | (0.006) | (0.012) |  |  |
| Born Between 1962–64*Famine Intensity |  |  | -0.004 | -0.009 |
|  |  |  | (0.006) | (0.013) |
| Other Controls | Yes | Yes | Yes | Yes |
| N | 1039 | 533 | 1672 | 1033 |

*Notes:* Data are from 2015-2017 Hurun Report. The unit of observation is an individual. *Wealth (Log)* is the natural log of the total amount of wealth held by individuals in the US dollars. *Famine Intensity* is the province-level average decrease in cohort size during the famine calculated from the 1990 China Population Census. *Born Between 1956-58* is a dummy variable for individuals born between 1956 and 1958. *Born Between 1962-64* is a dummy variable for individuals born between 1964 and 1964. Other control variables include birth year fixed effects, ranking year fixed effects and province of birth fixed effects. Robust standard errors adjusted for clustering within birth year. *** Significant at the 1% level. ** Significant at the 5% level. * Significant at the 10% level.

Table 5: Alternative Measure of Famine Intensity - All Sample

|  | Dependent Variable: Wealth (Log) | | |
|---|---|---|---|
|  | (1) | (2) | (3) |
| Born during the Famine* Excess Death Rate (1959–61) | -0.0187*** | -0.0186*** | -0.0189*** |
|  | (0.004) | (0.004) | (0.005) |
| Born before the Famine* Excess Death Rate (1959–61) | 0.004 | 0.004 | 0.004 |
|  | (0.008) | (0.008) | (0.008) |
| Birth Year FE | Yes | Yes | Yes |
| Province of Birth FE | No | No | Yes |
| Ranking Year FE | No | Yes | Yes |
| N | 2948 | 2948 | 2948 |

*Notes:* Data are from 2015-2017 Hurun Report. The unit of observation is an individual. *Wealth (Log)* is the natural log of the total amount of wealth held by individuals in the US dollars. *Excess Death Rate (1959-61)* is the province-level average excess death rate during the famine period calculated from the National Bureau of Statistics of China (1999). *Born during the Famine* is a dummy variable for individuals born between 1959 and 1961. *Born Before the Famine* is a dummy variable for individuals born before 1959. Columns 1 and 2 control for province level famine intensity but not shown here. Robust standard errors adjusted for clustering within birth year. *** Significant at the 1% level. ** Significant at the 5% level. * Significant at the 10% level.

# Online Appendix for

**The Persistent Effect of Famine on Present-Day China: Evidence from the Billionaires**

Pramod Kumar Sur

Asian Growth Research Institute (AGI) and Osaka University

Masaru Sasaki

Osaka University and IZA

Section A: Figures

This section presents the additional figures used in this paper. Figure A1 presents the birth and death rate in China between 1949 and 1990 per 1000 population. Figure A2 presents the province-level famine intensity measured by excess death rate. Figure A3 presents the average wealth of individuals based on their year of birth and average wealth by famine intensity of their province of birth in log US dollars. Figure A4 presents the event study analysis of the selection effect of famine in our sample. Figure A5 presents the event study analysis of the impact of exposure to famine on wealth for the whole sample considered in this paper.

Figure A1: Birth rate and Death rate in China 1949-1990 (Per 1000 population)

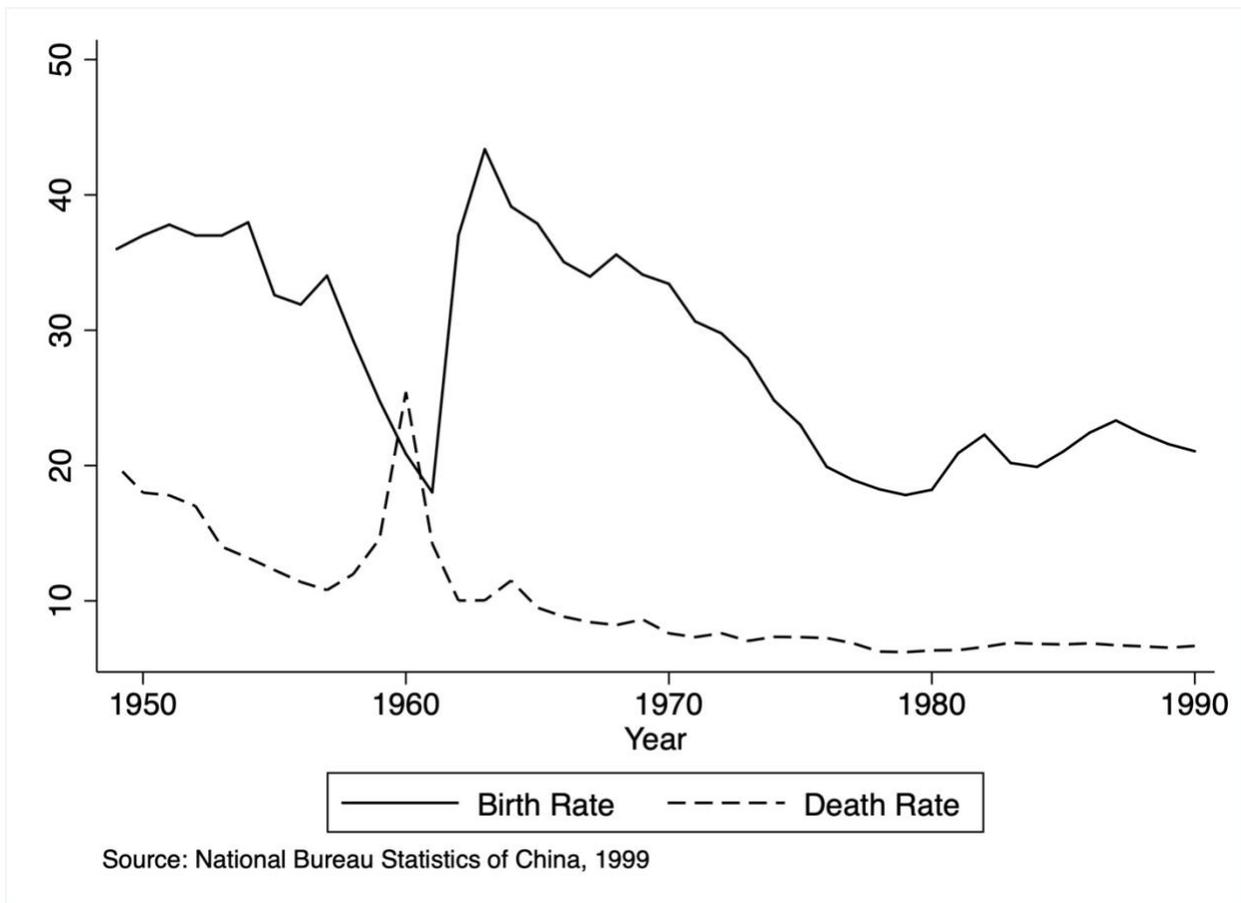

*Notes:* The figure presents the birth rate and death rate in China between 1949-1990. The black line plots the birth rate in china per 1000 population. The dotted line plots the death rate in China per 1000 population.

Figure A2: Province-level Famine Intensity - Excess Death Rate

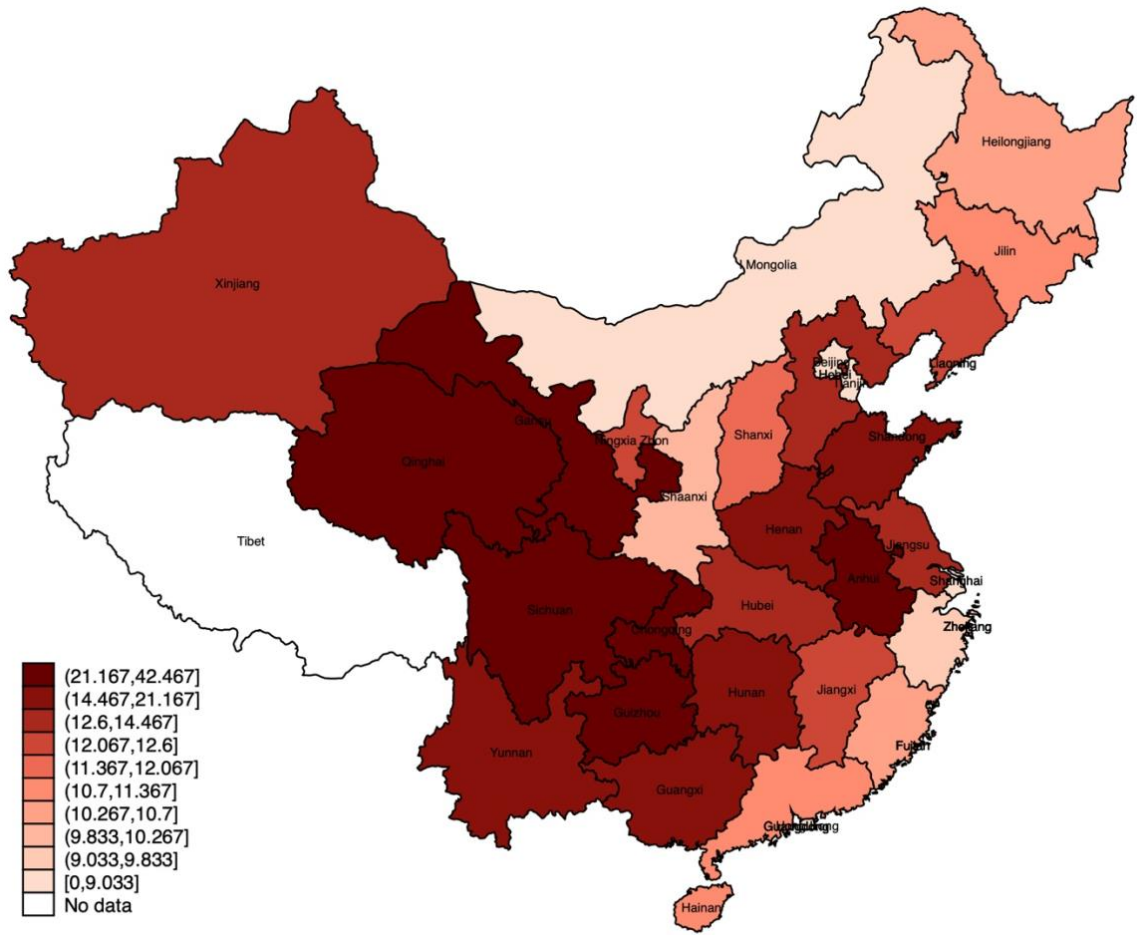

*Notes:* The figure presents the province-level variation in excess death rates measured by the percentage increase in death during the famine period in comparison with three years prior (i.e., 1956–58) and after (i.e., 1962–64) the famine period. Darker shades indicate larger share of excess death during the famine.

Figure A3. Average Wealth by Year of Birth (in log US dollars)

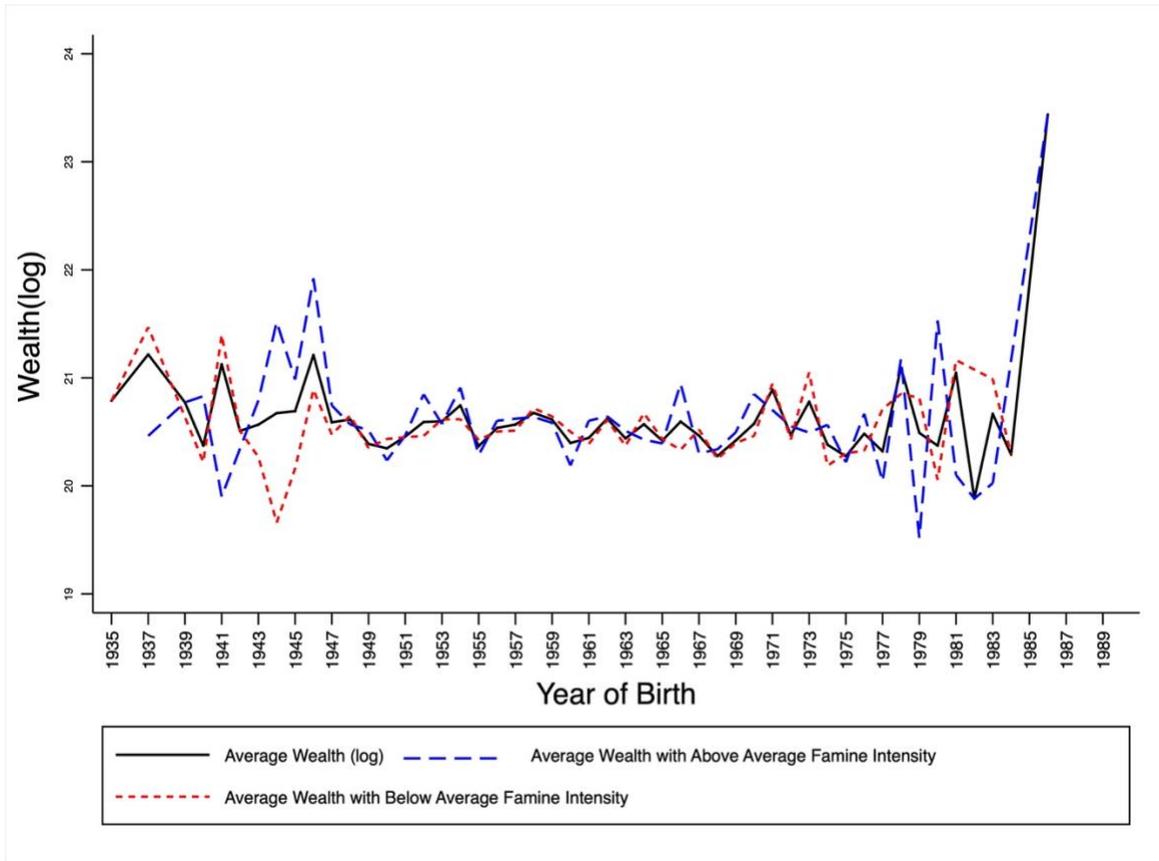

*Notes:* The figure reports the average wealth of individuals based on the year of birth and average wealth by famine intensity of the province of birth. The black line represents the average wealth of individuals by the year of birth. The blue and red lines represent the average wealth of individuals born in provinces where the famine intensity was above and below the average, respectively.

Figure A4. Selection Effect - Event Study Graph

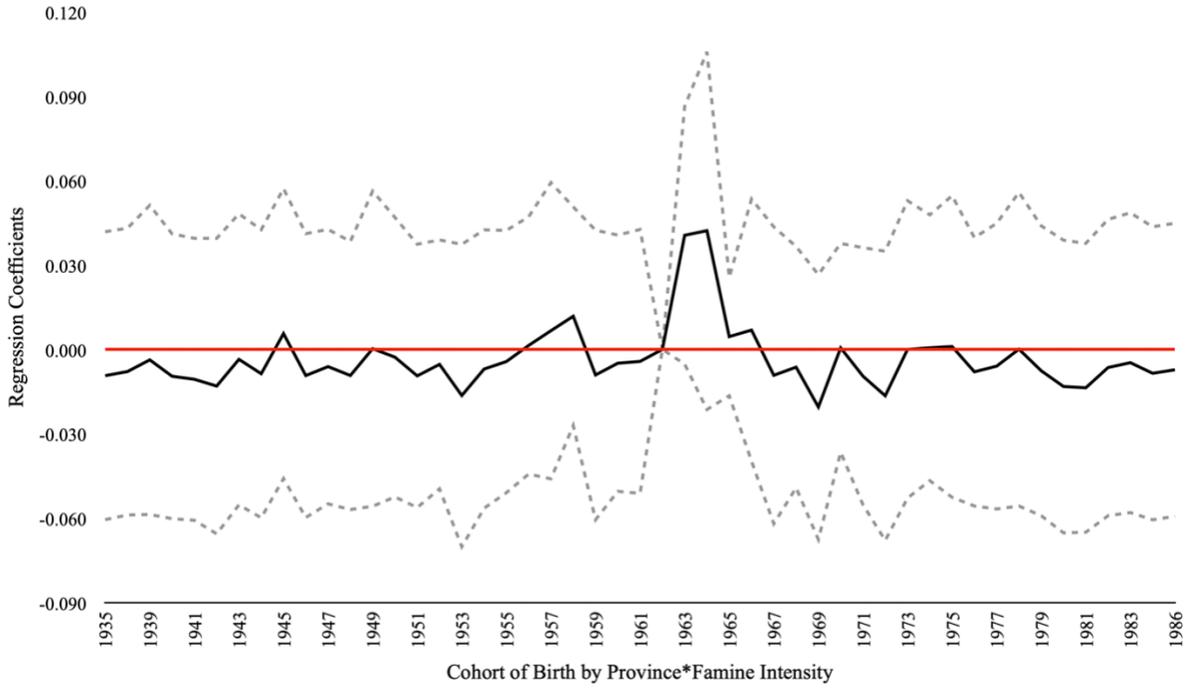

*Notes:* The figure presents the regression coefficients of the year-by-year effect of the selection effect of famine through event study graphs for the whole period. The black line reports the main impact with a 95% confidence interval presented in the dotted grey line.

Figure A5. Event Study - All Years

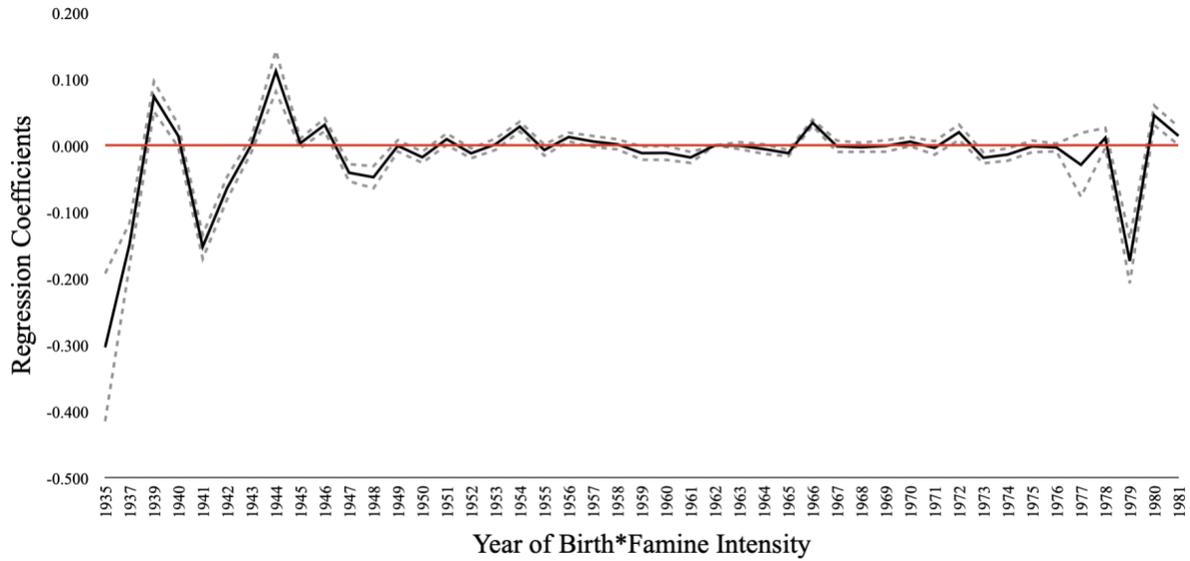

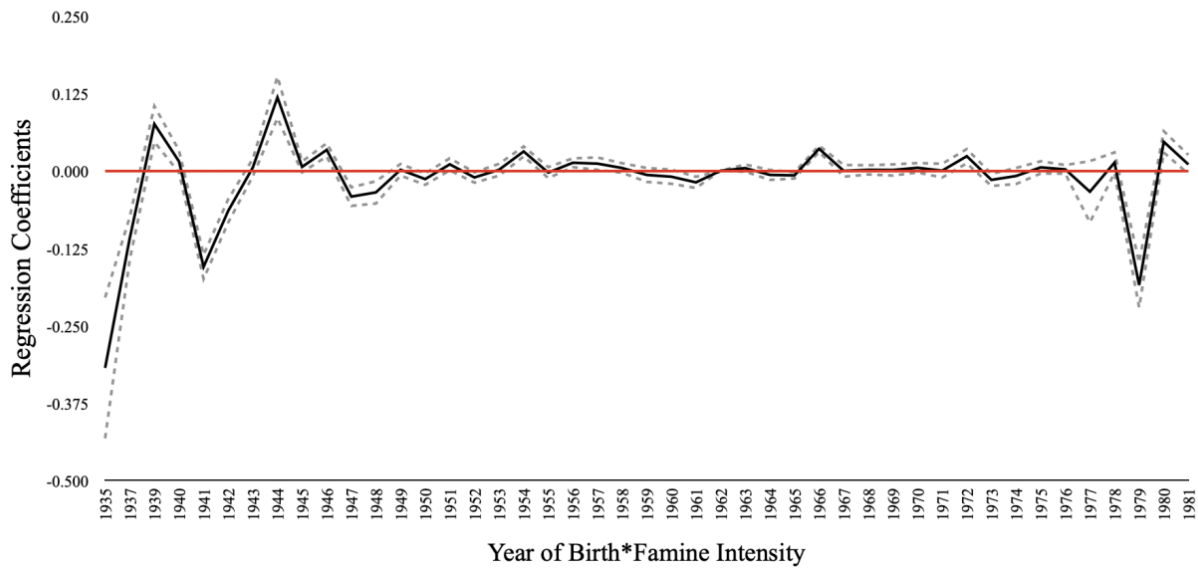

*Notes:* The figure presents the regression coefficients of the year-by-year effect of the exposure to the famine through event study graphs for the whole period. Panel A reports the event study graph for the whole sample (2015-17). Panel B reports the event study graph for 2016 sample only. The black line reports the main impact with a 95% confidence interval presented in the dotted grey line.

# Section B: Tables

This section presents the additional tables used in this paper. Table B1 presents the details about the construction of our sample data set for this paper. Table B2 presents the results on the impact of famine on cohorts of birth by province for sample excluding zero. Table B3 presents the impact of exposure to the famine on wealth, considering the 2016 sample only. Table B4 presents the results for section 4.3, considering alternative choices of clustering standard errors. Table B5 presents the robustness results for section 4.5, considering an alternative measure of famine intensity measured by excess death rate for 2016 sample only.

Table B1: Sample Construction

|  | Full Sample | Sample of individuals where birth year is available | Samples where province of birth is available | Sample considered in our analysis |
|---|---|---|---|---|
|  | (1) | (2) | (3) | (4) |
| Wealth (log) | 20.419 | 20.439 | 20.536 | 20.537 |
|  | (0.772) | (0.783) | (0.809) | (0.811) |
| Born before the Famine (Before 1959) | 0.219 | 0.301 | 0.274 | 0.352 |
|  | (0.414) | (0.459) | (0.446) | (0.478) |
| Born During the Famine (1959-61) | 0.084 | 0.084 | 0.080 | 0.080 |
|  | (0.278) | (0.278) | (0.272) | (0.272) |
| Famine Intensity | 37.291 | 37.278 | 37.291 | 37.278 |
|  | (10.396) | (10.540) | (10.396) | (10.540) |
| Birth Year | 1962.273 | 1962.273 | 1961.156 | 1961.156 |
|  | (8.513) | (8.513) | (8.296) | (8.296) |
| Sample Year | 2016.038 | 2016.026 | 2015.995 | 2015.994 |
|  | (0.812) | (0.812) | (0.803) | (0.803) |
| Maximum no of Observations | 6058 | 4415 | 3788 | 2948 |

*Notes:* Data are from the 2015-2017 Hurun Report. The table reports the mean and Standard deviations (in parentheses) of the data considered in this paper. Summary statistics are constructed based on the available data. Column 1 reports the details of the entire sample. Column 2 reports the details of the sample where the birth year data is available. Column 3 reports the details of the sample where the province of birth data is available. Column 4 reports the details of the samples considered in our analysis.

Table B2: Impact of Famine on Cohort of Birth by Province - Sample excluding zero

|  | Dependent Variable: Cohort Size | | |
| --- | --- | --- | --- |
|  | All Years | 1940-1980 | 1950-1970 |
|  | (1) | (2) | (3) |
| Famine Intensity*Famine Period | -0.024 | -0.024 | -0.028 |
|  | (0.015) | (0.016) | (0.019) |
| Famine Intensity*Before the Famine Period | -0.017 | -0.018 | -0.023 |
|  | (0.011) | (0.011) | (0.014) |
| Province FE | YES | YES | YES |
| Year FE | YES | YES | YES |
| Observations | 464 | 446 | 325 |

*Notes:* Data are constructed from 2016 Hurun Report. The unit of observation is a province. *Cohort Size* is the number of unique individuals born in each province each year in our sample. *Famine Intensity* is the province-level average decrease in cohort size during the famine calculated from the 1990 China Population Census. *Famine Period* is the year between 1959 and 1961. *Before the Famine Period* is the year before 1959. Robust standard errors adjusted for clustering within province. *** Significant at the 1% level. ** Significant at the 5% level. * Significant at the 10% level.

Table B3: Impact of Famine on Wealth - 2016 Sample

|  | Dependent Variable: Wealth (Log) | |
| --- | --- | --- |
|  | (1) | (2) |
| Famine Intensity*Born during the Famine | -0.0124*** | -0.0142*** |
|  | (0.004) | (0.004) |
| Famine Intensity*Born before the Famine | 0.004 | 0.003 |
|  | (0.005) | (0.005) |
| Birth Year FE | Yes | Yes |
| Province of Birth FE | No | Yes |
| N | 1049 | 1049 |

*Notes:* Data are from 2016 Hurun Report. The unit of observation is an individual. *Wealth (Log)* is the natural log of the total amount of wealth held by individuals in the US dollars. *Famine Intensity* is the province-level average decrease in cohort size during the famine calculated from the 1990 China Population Census. *Born during the Famine* is a dummy variable for individuals born between 1959 and 1961. *Born before the Famine* is a dummy variable for individuals born before 1959. Columns 1 control for province-level famine intensity but are not shown here. Robust standard errors adjusted for clustering within the birth year. *** Significant at the 1% level. ** Significant at the 5% level. * Significant at the 10% level.

Table B4: Alternative Choices of Clustering Standard Error

| | Dependent Variable: Wealth (Log) | | |
|---|---|---|---|
| | Robust Standard Error | Clustering at Birth Province Level | Two Way Clustering at Within Birth year and Within Province of Birth Level |
| | (1) | (2) | (3) |
| Famine Intensity*Born during the Famine | -0.0142** | -0.0142** | -0.0142*** |
| | (0.007) | (0.006) | (0.005) |
| Famine Intensity*Born before the Famine | 0.002 | 0.002 | 0.002 |
| | (0.005) | (0.005) | (0.005) |
| Birth Year FE | Yes | Yes | Yes |
| Province of Birth FE | Yes | Yes | Yes |
| Ranking Year FE | Yes | Yes | Yes |
| N | 2948 | 2948 | 2948 |

*Notes:* Data are from 2015-2017 Hurun Report. The unit of observation is an individual. *Wealth (Log)* is the natural log of the total amount of wealth held by individuals in the US dollars. *Famine Intensity* is the province-level average decrease in cohort size during the famine calculated from the 1990 China Population Census. *Born during the Famine* is a dummy variable for individuals born between 1959 and 1961. *Born before the Famine* is a dummy variable for individuals born before 1959. Column 1 reports robust standard errors without clustering. Column 2 reports standard errors adjusted for clustering within birth province level. Column 3 reports standard errors adjusted for two-way clustering within birth province and within birth year. *** Significant at the 1% level. ** Significant at the 5% level. * Significant at the 10% level.

Table B5: Alternative Measure of Famine Intensity - 2016 Sample

|  | Dependent Variable: Wealth (Log) | |
|---|---|---|
|  | (1) | (2) |
| Born during the Famine* Excess Death Rate (1959–61) | -0.0192*** | -0.0196*** |
|  | (0.004) | -0.006 |
| Born before the Famine* Excess Death Rate (1959–61) | 0.005 | 0.004 |
|  | (0.008) | (0.008) |
| Birth Year FE | Yes | Yes |
| Province of Birth FE | No | Yes |
| N | 1049 | 1049 |

*Notes:* Data are from 2016 Hurun Report. The unit of observation is an individual. *Wealth (Log)* is the natural log of the total amount of wealth held by individuals in the US dollars. *Excess Death Rate (1959-61)* is the province-level average excess death rate during the famine period calculated from the National Bureau of Statistics of China (1999). *Born during the Famine* is a dummy variable for individuals born between 1959 and 1961. *Born Before the Famine* is a dummy variable for individuals born before 1959. Column 1 controls for province level famine intensity but not shown here. Robust standard errors adjusted for clustering within birth year. *** Significant at the 1% level. ** Significant at the 5% level. * Significant at the 10% level.